\documentclass[12pt,preprint]{aastex}
\usepackage{rotating}
\begin{document}
\title{$\Lambda$CDM cosmology: how much suppression of credible evidence,
and does the model really lead its competitors, using all evidence?}

\author{Richard Lieu$\,^{1}$}

\affil{\(^{\scriptstyle 1} \){Department of Physics, University of Alabama,
Huntsville, AL 35899.}\\}

\begin{abstract}
Astronomy can never be a hard core
physics discipline, because the Universe offers no 
control experiment, i.e. with no
independent checks it is
bound to be highly ambiguous and degenerate.  Thus e.g.
while superluminal
motion can be explained by Special Relativity.  data on the former
can never on their own be used to establish the latter.  This is why
traditionally astrophysicists have been content with (and
proud of) their ability
to use 
known physical laws and
processes established in the laboratory to explain
celestial phenomena.  Cosmology is not even astrophysics: all
the principal assumptions in this field are unverified (or unverifiable)
in the laboratory, and
researchers are quite comfortable with
inventing unknowns
to explain the unknown.
How then
could, after
fifty years of failed attempt in finding dark matter, the
fields of dark matter {\it and now} dark energy have become such
lofty priorities in astronomy funding, to the detriment of all other branches
of astronomy?  I demonstrate in this article that while some of  is based
upon truth, at least just as much of $\Lambda$CDM 
cosmology has been propped by a paralyzing amount of
propaganda which suppress counter evidence and subdue competing models.
The recent WMAP3 paper of Spergel et al (2007) will be used as case in
point on selective citation.  
I also show that when all evidence are taken into account,
two of the competing models that abolish dark energy and/or
dark matter do not trail behind $\Lambda$CDM by much.
Given all of the above, I believe  astronomy is no longer
heading towards a healthy future, unless funding agencies re-think
their master plans by backing away from such high a emphasis on groping
in the dark.

\end{abstract}

\noindent
{\bf 1.  Introduction: on the shoulder of giants}

The history of science  is full of examples of
major breakthroughs being made by radical thinkers, those
who tend to ask silly questions and `chain their
coffee mugs to radiator pipes'.  No doubt, the rise
and fall of great scientific hypotheses are always brought about by the
availability of crucial new data, and if credit is truly given to
every source where credit is due, then the engineer who serviced one's
departmental Xerox machine should also be included as part of the
`team' which made a miracle possible.  But what provides
the shoulder upon which a theoretical giant stands, is usually
the ideas behind the
design of an
important experiment, or new and revealing
ways of approaching an old dataset.
These invariably come from a small number of highly creative minds.
I therefore begin this essay by voicing my unreserved support to
the courageous gesture of Simon White in his recent article 
`On why dark energy is bad for astronomy'
(White 2007), and by expressing open disagreement with the comments
of my great and long time Imperial College friend
Matt Mountain, who said  (Nature 2007, 447, 122) that cosmology in the
present era may no longer be driven by the chosen few.  Matt's
implication here, it would seem, is that this field
is now so special  and priviledged
that whatever we learned from the history of {\it any}
branch of science has no bearing on it, and can be disrespected.

\noindent
{\bf 2. Contemporary cosmology: rising above the shoulder of giants}

\noindent
It is not my intention to go through the past reminiscences of
$\Lambda$CDM - not even the near-term past.  I
wish to make just one point, viz. that the version of the model
as we often hear about it today has not been around for much longer
than a decade.  It was only in the early 90's, when I invited
the late Prof. David Schramm (Chicago) to deliver a lecture at
UC Berkeley on the age of the Universe, that he showed
a plot of the 
time evolution of the `most
favored' value of the cosmological
constant.  His graph looked like a sine curve with an average of zero.
Thus there is not so much scope for arguing that cosmologists
expected $\Lambda$CDM, less so predicting it.

Has $\Lambda$CDM cosmology `transcended' the scientific method?
Before we seek to answer the
question, I first wish to emphasize that
I believe in astronomy we should be
particularly proud of our own tradition, and should from time
to time ask this same question without feeling
insulted and irrespective of our own sub-field
of interest, {\it because} it was the astronomers who brought 
the world into the era of `modern' or `enlightenment' science, i.e.
we should set a {\it stricter} standard for ourselves.
It all began with what happens
up there - the heavenly motion - which could suddenly be 
explained in terms of what
happens down here - what keeps our feet on the ground.
Newton acquired the
status of the `Father of modern science' 
by his ability to use the `known' to demystify the
`unknown' (hence to dispel the `fear of the unknown' that
haunted the dark ages with myths and superstition).
The history of astronomy since Newton is filled with glorious
stories of how the unfamimliar phenomena found in `remote' corners
of the Universe could be reduced to something directly
or closely related to experiences in
our daily environment.
In Table 1 I provided a few recent examples of such
revelations which most of us feel
at home with.  These are to be contrasted with recent developments
in cosmology, in which {\it every} key observational result was
`explained'  (and rewarded with some of the most prestigious 
prizes of our times)
by postulating completely new physics which
received no laboratory verification.   Are we effectively endorsing
the critics of astronomy funding (our {\it real}
enemies), who advocate that we can
afford to play such games because `so little is at stake' in
cosmology?

Perhaps one should nevertheless pause and ponder.  Charging
under the banner of Einstein's extreme eminence and his
forbidding theory of General Relativity,
have cosmologists been
over-exercising our priviledges?  Should all of us
succumb to Matt Mountain's Nature testimony 
that `times are changing, abandon traditions and get used to
it'?   Indeed, even among the examples of Table 1 there are cases
in which astronomers helped to bolster (strengthen) the claims
and assumptions of physicists, i.e. they serve more than the purpose
of merely dispelling ignorant
superstitions about unsual celestial manifestations.
Thus e.g. a white dwarf star
provided the unique battleground, unachievable in the laboratory, 
in which
two separately well tested laboratory phenomena of electron
quantum degeneracy and proton classical gravity compete with
each other.  Given that none of the other elementary particle
interactions have been unified with gravity, it is very important
and re-assuring to know from this piece of astrophysics that at
the average inter-particle distance of white dwarf matter quantum
and classical effects live their own existences and can even
play tug-of-war.  Another example (not in Table 1) which pushes
the argument even further
concerns solar neutrinos.  Here, the discrepancy in numbers between
Bahcall's prediction and observations was at least in part responsible
for the eventual discovery of neutrino oscillations in the physics
community.

Yet the above interesting cases of `feedback'
from astronomy to physics means the only good news for astronomers (including
cosmologists) is that we could occasionally be employable to help do some
real
physics.  There is no further room for loftiness:
we have {\it not} seen
or heard a single story in which a drastic shift in our understanding
of the physical world was initiated and confirmed using
astronomical data {\it alone}, and the reason (which has nothing to do
with tradition) is as follows.
Whether the subject matter is as fundamental
as time and space, or something more mundane, astronomical observations
can never by themselves be used prove `beyond
reasonable doubt' a physical theory.  This
is because we live in only one Universe - the indispensable `control
experiment' is not available.  There is no possibility of `flipping this
switch' or `turning that knob'.  Using SN1a data alone, one
will never clinch $\Lambda$CDM to the level of rigor of Maxwell's
equations, because there will always be the niggling doubt, however
little, of whether SN1a are standard candles.   Large samples of data
always help, e.g. to probe foreground matter using weak lensing shear
distortions, one needs to observe no more than a few background
quasars if the intrinsic shape of each is known; but without this
`control' we can only rely on the statistical behavior of many quasars.
Alternatively, `cross checks' by merging together many diverse and
independently acquired datasets,  a favorite approach of P.J.E, Peebles,
will also serve the purpose.  Yet in cosmology both such efforts
inevitably lead to a exponential proliferation of costs, and in the end
neither will ever replace the simple action of `flipping the switch'.
Hence the promise of using the Universe as a physics laboratory from
which new incorruptible physical laws may be
established without the support of 
laboratory experiments is preposterous (Matt, perhaps
you should get used
to {\it this}?).
\begin{table}
\begin{center}
\small
\begin{tabular}{|l|c|c|c|}
\hline
\textbf{Phenomenon} & \textbf{Explanation} & \textbf{Seminal Paper} & \textbf{Based on Laboratory} \\
& & & \textbf{Established Physics?} \\
\hline
\hline
Planetary orbits & Universal gravitation & Newton & Yes\\
\hline
Tides & Universal gravitation & Newton & Yes\\
\hline
X-ray Bursts & Thermonuclear Flashes & Woosley, Taam & Yes\\
\hline
Her X-1 & Accretion & Hayakawa, Matsuoka, & Yes/Maybe\\
& &  Prendergast, Burbidge & \\
\hline
Superluminal Motion & Special Relativity & Martin, Rees, & Yes\\
& & Albert Einstein & \\
\hline
White Dwarf Star & Quantum Physics meets & Chandrasekhar & Known physics\\
& Gravity  & & individually verified\\
\hline
\end{tabular}
\caption{Examples of recent achievements of astrophysicists in re-assuring
mankind that unusual phenomena in the sky do not have to mean bad omen:
they can be explained in terms of the physical laws here on earth that
we are familiar with.}
\end{center}
\end{table}

\vspace{1cm}

\begin{table}
\begin{center}
\small
\begin{tabular}{|l|c|c|c|}
\hline
\textbf{Phenomenon} & \textbf{Explanation} & \textbf{Based on laboratory} & \textbf{Verifiable in}\\
& & \textbf{establised physics?} & \textbf{future experiments?}\\
\hline
\hline
Redhsift & Expansion of Space & No & Unverifiable \\
& & & $\big($ Chodorowski 2007 $\big)$ \\
\hline
CMB & Big Bang & No & Far Future \\
\hline
Rotation Curves & Dark Matter & No & Near Future \\
& & & $\big($ As always $\big)$ \\
\hline
Distant Supernovae & Dark Energy & No & Far Future \\
\hline
Flatness and Isotropy & Inflation & No & Remote Future\\
\hline
\end{tabular}
\caption{Cosmologists {\it only know how to use} 
`unknowns' to explain `unknowns' (and
hence are not really astrophysicists).  In mainstream physics
new postulates are sometimes made to help account for unexpected
phenomena found in the laboratory, but the Universe is not
a laboratory because one crucial fundamental criterion:
the need for control experiments, cannot be met.}
\end{center}
\end{table}

\noindent
{\bf 3.  Cosmic microwave background (CMB):
a clean and direct test of its origin?}

I now venture the `how dare you' question, on the origin of the CMB
(and always bearing in mind that Feynman must find his own way of
convincing himself).  How do we know that it is the afterglow of the
Big Bang?  Here are some of the ususal responses I heard of, since
I was a school boy some 30 years ago.

\noindent
$\bullet$ {\it It comes from a redshift of}  1,000.  How do you know
the redshift of the CMB?  We do not have any characteristic emission
or absorption line - there is not even a straw to clutch.

\noindent
$\bullet$ {\it It comes from every direction in the sky, and is extremely
uniform}.  Well?   I recall the days when BATSE aboard CGRO discovered
that gamma-ray bursts (GRBs) are isotropically distributed. Yet, unlike
the cosmologists, the GRB community seems to be more careful in going
about the conjecture.  There was a Great Debate at the Shapley-Curtis
level, between Paczynski and Lamb, on how much can one make of the
distance scale of GRBs from the isotropy of the source distribution.
Cosmologists took a shortcut - the conclusion was drawn without a
Great Debate.  Since the CMB is more fundamental than GRBs,
as I expect my colleagues in the GRB community would probably agree,
why a Great Debate never took place on this subject?

\noindent
$\bullet$ {\it The CMB spectrum is a perfect black body, pointing clearly
and unequivocally back to the era of strong coupling between matter
and radiation}.  Yet how much does this constitute a {\it proof}, as opposed
to mother nature laughing at us having completely
missed some physical process that takes place in e.g. empty space?

\noindent
$\bullet$ {\it The CMB temperature was predicted by Big Bang theoreticians.}
Well, the prediction by Gamov was off by an order of magnitude.
Where do you draw the line?  How would you like the temperature of
your room be increased ten-fold?

\noindent
$\bullet$ {\it There are tiny temperature anisotropies in the CMB that can
beautifully be explained in terms of Big Bang cosmology}.  This is
provided one assumes the anisotropies are also cosmological in origin,
and that the early Universe comprised dark matter, dark energy, and
underwent a mysterious epoch of `inflation' to secure a delicate
balance of proportions between the two `dark' components.   Can
we frankly say that this is a {\it clean and straightforward} proof,
when so many other strings and loose ends are attached?

\noindent
$\bullet$  {\it How about all of the above?  Do you have a better
interpretation of the CMB?}  In no reasonable court of law will a
suspect be convicted of murder simply because there have been no other
arrests or suspects.

It is clear that all of the aforementioned arguments, even taken
together, only constistute an incoherent collection of circumstantial
evidence.  To make cosmologists worthy of the billions of governmental
support, we actually need to do {\it much} better than  convicting
a suspect of murder.  At best it would seem that we achieved what a
Scottish jury would return as verdict: not proven.

I must then turn to a different question.  Is there any direct way
of clinching the origin of the CMB at all that we can
pursue?  Well, it turns out there {\it is} one good starting
point.  Traditionally, just about the only clean and
indisputable way of charting the
scale height of any diffuse radiation in the absence of
redshift information is to look for `shadow' effects on the radiation
cast by gas clouds at known distances away from us: if
a shadow is found, the radiation must have come from {\it behind} the
cloud.  Can I provide n example to dramatize this point?  The
answer is also yes, except the outcome did present a major surprise to the
researchers in the field of concern.  It has for
a long time been generally accepted that the soft X-ray sky background (SXRB)
is principally Galactic in origin; moreover it is emitted by a thin hot
plasma that fills a void of $\sim$ 100 pc radius centered at the sun - the
so-called `local bubble' model.  Thus, when the first shadow of the
SXRB was discovered by
Burrows \& Mendenhall 1991, who reported
that 60 \% of the SXRB was silhouetted by a dense cloud - the Draco Nebula 
located some 600 pc away in a direction of high Galactic latitude - the
`local bubble' model was under serious threat.  Today, this great discovery
still stands, and completely destroyed any simple way of understanding
the SXRB origin.  Could there be a lesson here for the unerring community
of $\Lambda$ CDMcosmologists?

For the CMB an equivalent `silhouette' cloud would be a cluster of galaxies,
which Thomson scatters CMB photons on the Rayleigh-Jeans part of the
CMB spectrum - the Sunyaev-Zel'dovich effect (SZE).
A recent study of $\sim$ 100 rich clusters, using WMAP
W band data (Bielby \& Shanks 2007) confirmed our earlier findings from
a smaller sample of 31 rich clusters (Lieu et al
2007) that WMAP detected almost no SZE
at all from these clusters, i.e. the CMB appears to have failed the
`shadow test'.  In particular the analysis by Bielby \& Shanks of 38 
clusters at a mean redshift of $z \approx$ 0,3 revealed a level of SZE
statistically consistent with  a null effect and completely inconsistent
with the expectation (they also truncated the predicted SZE profile at
a ridiculously small cluster radius, yet the inconsistency remains, so that
its origin {\it cannot} be WMAP's spatial resolution).
When the same sample of 38 clusters was observed
by radio interferometric techniques rather than WMAP, a different verdict
was delivered.  This method did lead to the discovery of SZE at the
expected level (Bonamente et al 2006).  

Who is right?  It is usual to
settle such discrepancies by appealing to the community at large for
independent analysis of the same datasets.  For WMAP this is possible,
because the data are all in the public domain, which is how Bielby \& Shanks
were able to cross-check Lieu et al (2006).  The observations of 
Bonamente et al (2006) are however not public: it is apparently
quite normal for SZE data taken by ground-based telescopes
to remain inaccessible by the rest of us for a long time.  It is fair
to say that the release of data (in as primitive (or unprocessed) a form
as possible) of
important experiments for
everyone to check helps bolster the claims of the original researchers
who `creamed the crop', especially if others who did the necessary tests
are able to corroborate these  claims.

Let us give $\Lambda$CDM proponents the full benefit of the doubt.
by assuming that the interferometers got it right, viz. the SZE
at the fully expected level as reported by Bonamente et al (2006)
is correct.  This then would mean, unless both Bielby \& Shanks
and Lieu et al erred, that WMAP got it wrong.  For the
clusters analyzed by Bielby \& Shanks (2006) and Lieu et al (2007),
their SZE profiles
have typical angular sizes between those of the first and second
acoustic peaks, except of course the amplitude of the SZE is deeper
than that of the acoustic oscillations.  If WMAP could not properly
fathom those deeper modulations in the CMB temperature that occur
at $\sim$ 0.5 degree angular scales, how shall we satisfy ourselves
that it has correctly measured the acoustic peaks?

\vspace{3mm}

\noindent
{\bf 4.  $\Lambda$CDM cosmology: some of 
the long list of counter evidence and how they have been treated}

In Table 3 I listed some of the counter evidence of 
$\Lambda$CDM cosmology, all of which
were published (or about to be published) in the topmost 
astronomy journals.   The table entries are referred to as `neglected
evidence' because I used the latest WMAP3 cosmology paper of
Spergel et al (2007) as benchmark concerning the citations of relevant
previous work.  Not only is this the most important cosmology paper in the
contemporary literature (it already received more than 1,300 ADS citations
even though it is not yet published),  but also it included many sections
on CMB external correlations, viz. how the standard $\Lambda$CDM model
fares against other non-CMB observations, and whether these can help
to further constrain the model parameters.  Thus the paper {\it is} meant
to cross-compare all the vital evidence, yet with the exception of
the `Axis of evil', none of those listed in  Table 3 were mentioned.
In the last section Spergel et al concluded `the standard
model of cosmology has survived another rigorous set of tests'.
Here I elaborate upon most of items of Table 3 (though not necessarily in
the right order) so that each reader
can judge if Spergel's
claim is tenable.  

\clearpage

\begin{sidewaystable}
\scriptsize
\begin{tabular}{|l|l|l|}
\hline
& &\\
\textbf{$\Lambda$CDM : the neglected evidence} & \textbf{Why it is important} & \textbf{`Reason' for neglect}\\
& &\\
\hline
\hline
Evolution of cluster counts & Curve matches Einstein de Sitter Universe & Cluster `detailed physics' not \\
$\big($Vauclair et al.$\big)$ & 
and excludes $\Lambda$CDM at 7 $\sigma$ & known $\big($but we understand the \\
& & Universe$\big)$\\
\hline
Missing baryons at low redshift & Still not found today as & Who cares about a few percent of the \\
$\big($Cen and Ostriker$\big)$ & cluster OVII lines remain undetected & Universe $\big($except this is the only bit \\
& & we directly measure, and we are \\
& & made of it$\big)$\\
\hline
Too little Sunyaev - Zel$'$dovich effect & `Shadowing' techniques 
still the only direct & Cluster physics `details'.  WMAP \\
in WMAP $\big($Bielby and Shanks, Lieu et al.$\big)$ & way of 
charting the CMB scale height & cannot measure SZE anyway\\
& & $\big($but can probe the shallower acoustic \\
& & peaks at similar angular scales$\big)$\\
\hline
Matter Budget of galaxy groups & Many groups $\big($like our own Local Group$\big)$ & Groups of galaxies are not properly \\
$\big($Ramella et al.$\big)$ & could easily hold altogether $\Omega$ 
$\sim$ 1 & weighed and counted\\
& worth of matter &\\
\hline
Axis of evil, correlation with HI clouds & Significant foreground issues remain & Statistics unclear $\big($and the same \\
$\big($Land, Verschuur$\big)$ & down to acoustic peak scales & `Bayesian prior' that \\
& &established $\Lambda$CDM can be used \\
& & to marginalize these$\big)$ \\
\hline
Hubble constant of Sandage et al & Significantly different from Freedman's & 
Systematic problems\\
& value even though both used HST data &\\
\hline
Soft X-ray excess in clusters (Bonamente, & See Fig. 1 $\big($can $\Lambda$CDM explain this ?$\big)$ & Minor obscenity. Phenomenon doesn't \\
Nevalainen, Kaastra, Fabian, Lieu) & & really exist\\
\hline
Dwarf elliptical rotation curves & Data give constant density cores whereas &
Poor spatial resolution of data, and  no \\
& $\Lambda$CDM halo profiles have central cusps & independent M/L ratio
for disc \\
\hline
\end{tabular}
\begin{center}
\caption{A list of key, independent and respectable evidence not cited in the
WMAP3 paper of
Spergel et al (2007), where the authors included an extensive
section on CMB external correlations
to bolster their claim of the standard
$\Lambda$CDM cosmological model.  Their
ability to `bolster' is because the
external evidence employed were carefully pre-selected.}
\end{center}
\end{sidewaystable}

\noindent
$\bullet$ Spergel et al cited the gas fraction analysis by Steve Allen of 
X-ray (Chandra) observations of rich clusters, which led Allen to
conclude upon the correctness of $\Lambda$CDM cosmology.  Yet it is
well known at least within the clusters community that the number
density evolution curve of clusters as derived by the XMM
Newton Key Project (Vauclair et al 2003)
rejected the $\Lambda$CDM prediction with
7 $\sigma$  statistical significance, but is consistent with an
Einstein de Sitter Universe.  This result was not cited by
Spergel et al (2007).  I understand that there has been a lot of
questions about the validity of the conclusion of Vauclair et al (2003),
but just as many similar questions have also been directed at the
work of Allen.  It is for the moment not for any individual to pass
ultimate judgements.  Even if the WMAP team has their own noble
rationale to favor Allen's work, they should still have cited Vauclair
because the claim is so drastic.  Quoting the paper in negative light
might still be acceptable if at least some reasons are given.  Ignoring
the paper altogether is unacceptable and unscientific.

\noindent
$\bullet$ The fact that only $\approx$ 50 \% of the baryons predicted
by the $\Lambda$CDM model to exist at low
redshifts has been observed was first noted by Cen \& Ostriker (1999).
To date this same problem persisted, with the latest paper on the
subject being Takei et al (2007).  In Figure 2 we show the upper
limits to-date on
the detection of these missing baryons (graph is courtesy Yoh Takei),
resulting partly from the non-detection of the O VII line in clusters
with soft X-ray excess, i.e. this excess cannot then be attributed to
a massive component of warm baryons at the outskirts of clusters.
This to me is a very serious discrepancy, much more so than the
debate on dark matter and dark energy, because {\it baryons are real}
and they are still (however one may get fancy) the only thing we
can directly measure.

\noindent
$\bullet$ The soft X-ray excess of clusters has over the past twelve
years since its discovery been detected by EUVE, ROSAT, BeppoSAX, XMM,
and Suzaku, with the latest paper (on Suzaku's signal) being Werner
et al (2007).  There is still no explanation of this excess, which is
seen in both the core and outskirts of clusters, in the context of
$\Lambda$CDM or for that matter any other cosmologies.  For those
who prefer to sideline this as yet another minor detail, I invite them
to take a look at the strength of the soft X-ray signal in the
central (but avoiding the complicated innermost) region of Abell 3112,
Figure 2.

\begin{figure}[!h]
\begin{center}
\includegraphics[angle=0,width=4in]{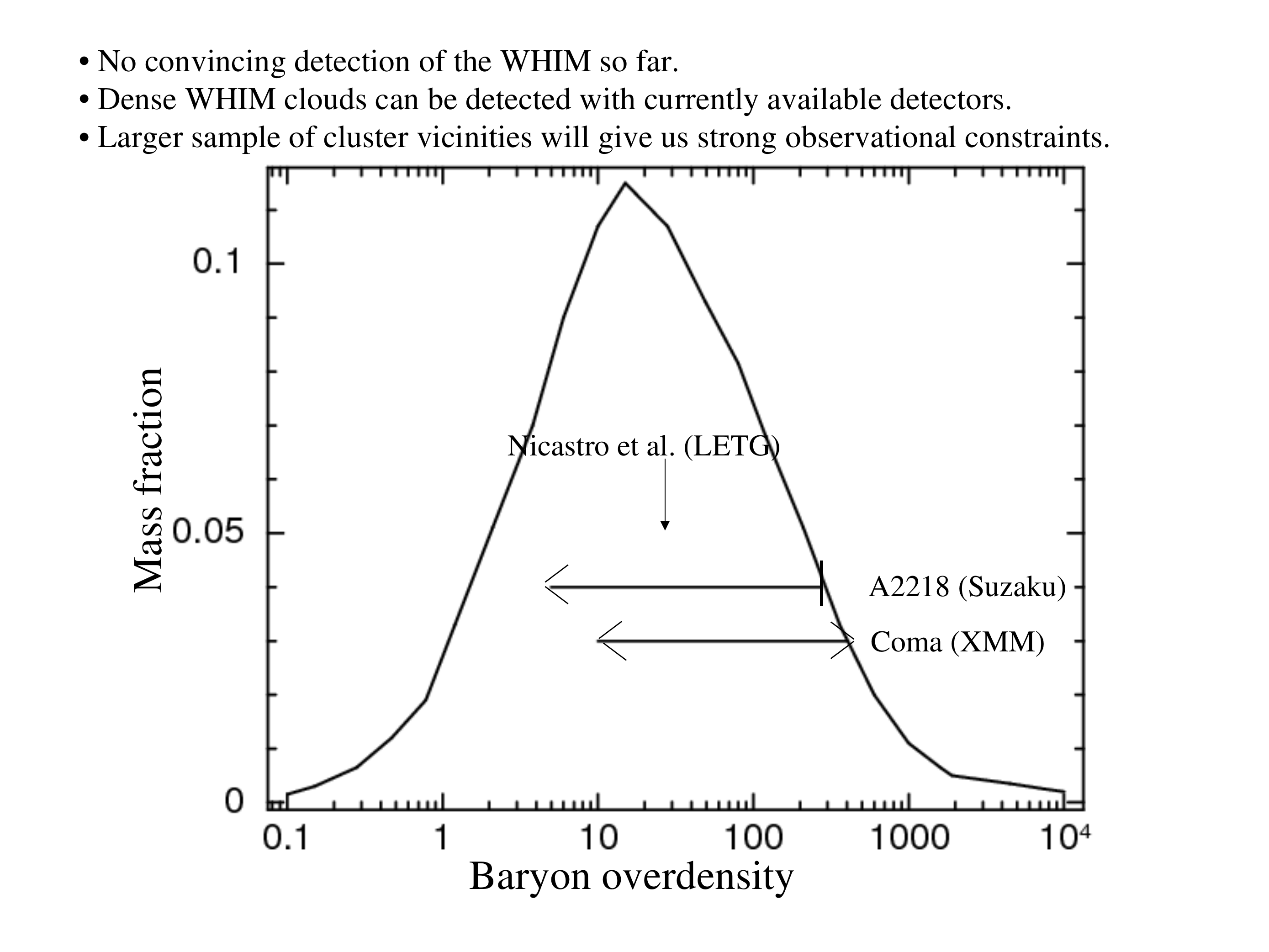}
\end{center}
\caption{Mass fraction of baryons (to total matter) as a function
of clump overdensity, with various upper limits and one 
3 $\sigma$ detection from Coma cluster.  Courtesy of Yoh Takei.}
\end{figure}

\noindent
$\bullet$ On the Hubble constant, Spergel et al (2007) cited the Hubble
Key Project paper of Freedman et al (2001) and the X-ray/SZE result
of Bonamente et al (2006), but the equally comprehensive treatise of
Sandage et al (2006) was ignored.  This could simply be due to the
very recent appearance of the Sandage, but since
in principle there is definitely enough time for
Spergel et al to cite Sandage there may be other reasons.
While Freedman and Bonamente reported a Hubble constant of $h \approx$ 0.7,
very close to the value advocated by the WMAP team, Sandage found
$h \approx$ 0.62, considerably lower perhaps than any team member's
liking.  Besides, how could two independent analyses of the
HST data (Freedman versus Sandage)  could lead to such a difference
in the final answer, especially since I have been hearing so
much talk about `$H_0$ in the era of precision cosmology is
nailed to 5 \% accuracy' ?

\noindent
$\bullet$  A not-too-often mentioned
but no less important problem for $\Lambda$CDM
is the potential for groups of galaxies like our own Local Group to
be harbors of much more matter than expected.  Thus e.g. from the ESO survey
of 1,168 nearby groups (Ramella et al 2002)  the mean virial mass per
group is $M \approx$ 1.15 $\times$ 10$^{14}$ M$_\odot$ and the number 
density of groups is $n \approx$ 1.56 $\times$ 10$^{-4}$ Mpc$^{-3}$.
This already yields a mean mass density $nM$  equivalent to
$\Omega_{{\rm groups}} \approx \Omega_m/2$, assuming the
$\Lambda$CDM value of $\Omega_m =$ 0.3.  However, there is a selection
bias, due to many groups having evaded detection.  After correcting
for this bias in the best possible way, Ramella et al (2002) estimated
a number density of $n \approx$ 4 $\times$ 10$^{-3}$ Mpc$^{-3}$ for
groups.   The product $nM$ now corresponds to 
$\Omega_{{\rm groups}} \approx$ 3.4, which far exceeds the total
mass density of matter in the $\Lambda$CDM model.  
The same pointers to groups of galaxies weighing much more
massively than `expectation' (i.e. $\Omega_m \approx$ 1) was also
found by Myers et al (2003, 2005).

\begin{figure}[!h]
\begin{center}
\includegraphics[angle=0,width=4in]{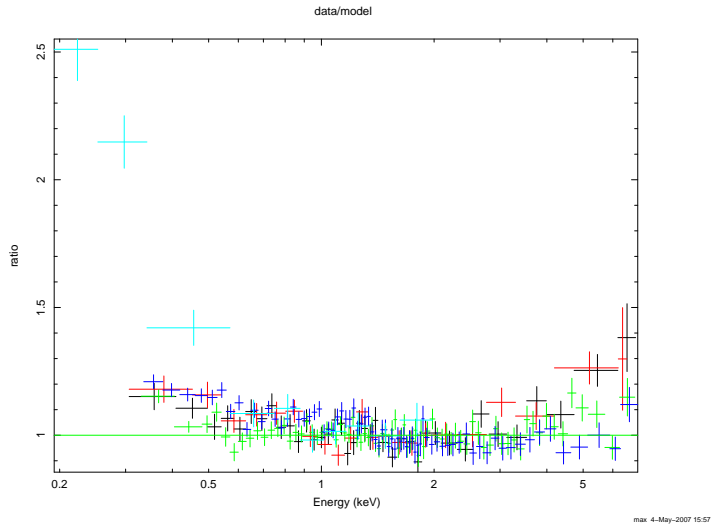}
\end{center}
\caption{Isothermal free-free emission model for the hot virialized
plasma in the rich cluster Abell 3112 as fitted to the X-ray
spectra of the central
0.5 -1.5 arcmin region of the cluster (the innermost 0.5 arcmin
was avoided due to possible complication from a `cooling core' and
point source contamination).  The graph shown plots the ratio between
observed data and the best-fit model, where the different colors
correspond to observations by the various X-ray missions: Chandra
in black and red, XMM in green and blue, and ROSAT PSPC in Cyan.
Note the strong soft X-ray excess at low energies, indicative of
a completely new emission component in clusters of galaxies.
This `cluster soft excess' phenomenon has been known and ridiculed
for twelve years, i.e. it lived a time span as long as $\Lambda$CDM.}
\end{figure}

\begin{figure}[!h]
\begin{center}
\includegraphics[angle=0,width=4in]{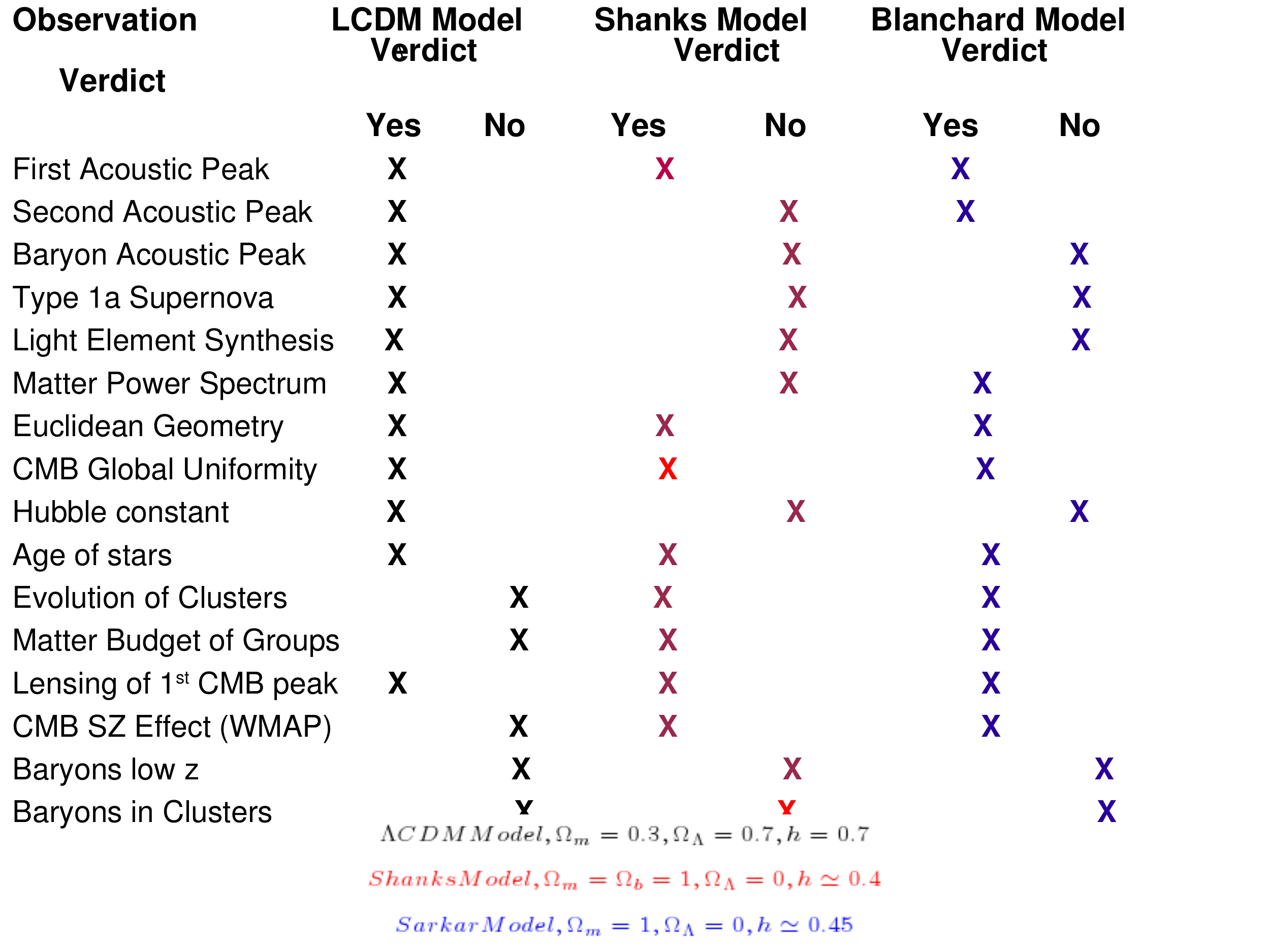}
\end{center}
\caption{How the standard $\Lambda$CDM model fares against two
competing models, those of Shanks (2007) and Blanchard-Sarkar et al
(2003), when all the evidence known to the author are taken into account.
Note the slight vertical misalignment of the crosses should be ignored -
the verdict from each bullet of evidence is for simplicity
only expressed as a binary yes or no.}
\end{figure}

\noindent
$\bullet$ The very feeble SZE detected by WMAP was already discussed
in the previous section, and here I simply mention that David Spergel is
fully aware of at least Lieu et al (2007).  It was not cited in
Spergel et al (2007).

\noindent
$\bullet$  There has recently been the claim by Verschuur et al (2007)
that a significant fraction of the degree-scale acoustic peak hot
spots in the ILC map of WMAP1 spatially correlates with anomalous
velocity HI clouds in the Milky Way - a finding which prompted this
author to conclude that a significant fraction of the WMAP 
anisotropy at the primary acoustic peak is not cosmological.
This paper was submitted to ApJ, and I was told that two reasonably
disposed
referee reports were received.   Readers should therefore keep an eye on
the development of this front.  No complaint is made in the present
article of Spergel's failure to cite Verschuur, as the latter is
a very new result.

\vspace{3mm}

\noindent
{\bf 5.  Alternative models: are they really so inferior to
$\Lambda$CDM?}

Given that there are so many bullets of evidence (with varying
weights) against $\Lambda$CDM cosmology, the question is naturally
raised as to how competing models may fare, when the whole body of
evidence is taken into account.  I show in Figure 3 just such a metric.
The two alternative models I chose are Shanks (2007) and Blanchard et al
(2003).  Both involve the Einstein de Sitter Universe, with
$\Omega_m =$ 1.  The latter does away with dark energy altogether,
and relies on a primordial matter spectrum that is not purely  power-law
(the level of extra contrivance here is not as severe as postulating
dark energy when dark matter has still not be found).   The former does
away with dark matter as well as dark energy, and uses the gravitational
lensing by foreground galaxy groups (see above) to secure agreement
between model prediction of the first acoustic peak and WMAP data.
Although the 2nd peak is not yet accounted for, the remarkable
feature of this Shanks model lies obviously with its economy in extra
new postulates, by getting rid of all darknesses.

It can be seen that when all the evidence are placed on the
`scale pan' no model is really classifiable as a `winner' or
`loser'.  Perhaps all models are equally poor:
the two competitors certainly do not come across as much
more inferior than the standard model.  What cannot be quantified
in terms of figure-of-merit, however, is how much more credibility
should one assign to a model that relies on less bizarre postulates.

\newpage

\noindent
{\bf 6. Conclusion}

Cosmologists should not pretend to be mainstream physicists,
because there is only one irreproducible Universe  and  control
experiments are impossible.  The claim to overwhelming evidence
in support of dark energy and dark matter is an act of exaggeration
which involves heavy selection of evidence and an inconsiderate attitude
towards alternative models with fewer (or no) dark components.
When all evidence are taken into account, it is by no means clear
that $\Lambda$CDM wins by such leaps and bounds.

Thus I do not see the wisdom of funding agencies in planning
such ambitious and expensive programs to perform dark energy research,
to the detriment of other fields of astronomy, as though cosmology
has now become a branch of physics, which it will never be.
These programs all have the common starting point that dark energy
is really out there - no question about it.  I hope the present
article demonstrated the contrary. 

The irony of today's times is that while dark matter is still
unidentified despite half a century of search, taxpayers are asked 
to invest in yet another potential fiasco.  Furthermore, the
situation as it evolves in time is that the more we do not find
dark matter, the less (in relative funding) do we invest in 
alternative approaches - to the point of totally choking these
approaches.
Thus we are putting more and more eggs in a less and less likely
basket.  Could this be the sign of a person (or  a
camp of people in prestigious institutes) who became angry because
they are embarrassed?  Even if one were to avoid taking such a view,
one should still ask the question `is this the scientific method' ?

I recommend that major funding agencies serious consider enlisting
to decision making panels  a higher (than zero) fraction of
those individuals who published equally respectable papers in
top journals on the body of counter evidence listed in Table 3.
The reason why we are heading in such wrong directions is because
while panels rotate they invariably comprise the same camp of
researchers, mostly from elite establishments with vested interests.
The ultimate selection effect, therefore, might lie with those 
senior agency
administrators responsible for the composition of these panels.

\noindent
{\bf 7.  References}

\noindent
Bielby, R.M., \& Shanks, T. 2007, MNRAS submitted (astro-ph/0703470).

\noindent
Blanchard, A. et al 2003, A \& A, 412, 35.

\noindent
Bonamente, M. et al 2007, ApJ, 647, 25.

\noindent
Burrows , D.N. \& Mendenhall, J.A., 1991, Nature, 351, 629.

\noindent
Cen, R., \& Ostriker 1999, ApJ, 514, 1.

\noindent
Chodorowski, M. 2007, MNRAS in press (astro-ph/0610590).

\noindent
Freedman, W.L. et al 2001, ApJ, 553, 47.

\noindent
Lieu, R. et al 2006, ApJ, 648, 176.

\noindent
Myers, A.D. et al 2003, MNRAS, 342, 467.

\noindent
Myers, A.D. et al 2005, MNRAS, 359, 741.

\noindent
Ramella, M. et al 2002, AJ, 123, 2976.

\noindent
Sandage, A. et al 2006, ApJ, 653, 843.

\noindent
Shanks, T. 2007, MNRAS, 376, 173.

\noindent
Spergel, D. et al 2007, ApJ in press (astro-ph/0603449).

\noindent
Takei, Y. et al 2007, ApJ, 655, 831.

\noindent
Vauclair et al 2003, A \& A, 412, L37.

\noindent
Verschuur, G. et al 2007, MNRAS submitted (arXiv:0704.1125)

\noindent
Werner, N. et al 2007, A \& A in press (arXiv:0704.0475).

\noindent
White, S.D.M. 2007, Rep Prog Phys, in press (arXiv:0704.2291).

\end{document}